\PassOptionsToPackage{table}{xcolor}
\documentclass[aip,apl,amsmath,amssymb,reprint]{revtex4-1}

\usepackage{chemformula} 
\usepackage[T1]{fontenc} 
\usepackage{graphicx}  
\usepackage{float}
\usepackage{amsmath}
\usepackage{amssymb}
\usepackage{array}
\usepackage{graphicx}
\usepackage{esint}
\usepackage[hidelinks]{hyperref}
\usepackage{placeins}
\usepackage{color}
\usepackage{soul}
\definecolor{tablecolor}{HTML}{89a1dc}

\begin{document}

\title{Hexagonal boron nitride cavity optomechanics}

\author{Prasoon K.\ Shandilya}
\thanks{These two authors contributed equally}
\affiliation{Institute for Quantum Science and Technology, University of Calgary, Calgary, AB, T2N 1N4, Canada}
\author{Johannes E.\ Fr{\"o}ch}
\thanks{These two authors contributed equally}
\affiliation{Institute of Biomedical Materials and Devices, University of Technology Sydney, Ultimo, NSW (2007) Australia}
\author{Matthew Mitchell}
\affiliation{Institute for Quantum Science and Technology, University of Calgary, Calgary, AB, T2N 1N4, Canada}
\author{David P.\ Lake}
\affiliation{Institute for Quantum Science and Technology, University of Calgary, Calgary, AB, T2N 1N4, Canada}
\author{Sejeong Kim}
\affiliation{Institute of Biomedical Materials and Devices, University of Technology Sydney, Ultimo, NSW (2007) Australia}
\author{Milos Toth}
\affiliation{Institute of Biomedical Materials and Devices, University of Technology Sydney, Ultimo, NSW (2007) Australia}
\author{Bishnupada Behera}
\affiliation{Institute for Quantum Science and Technology, University of Calgary, Calgary, AB, T2N 1N4, Canada}
\author{Chris Healey}
\affiliation{Institute for Quantum Science and Technology, University of Calgary, Calgary, AB, T2N 1N4, Canada}
\author{Igor Aharonovich}
\affiliation{Institute of Biomedical Materials and Devices, University of Technology Sydney, Ultimo, NSW (2007) Australia}
\author{Paul E.\ Barclay}
\affiliation{Institute for Quantum Science and Technology, University of Calgary, Calgary, AB, T2N 1N4, Canada}
\email{pbarclay@ucalgary.ca}

\begin{abstract}
Hexagonal boron nitride (hBN) is an emerging layered material that plays a key role in a variety of two-dimensional devices, and has potential applications in nanophotonics and nanomechanics. Here, we demonstrate the first cavity optomechanical system incorporating hBN. Nanomechanical resonators consisting of hBN beams with predicted thickness between 8 nm and 51 nm were fabricated using electron beam induced etching and positioned in the optical nearfield of silicon microdisk cavities. A 160 fm/$\sqrt{\text{Hz}}$ sensitivity to the hBN nanobeam motion is demonstrated, allowing observation of thermally driven mechanical resonances with frequencies between 1 and 23 MHz, and mechanical quality factors reaching 1100 at room temperature in high vacuum. In addition, the role of air damping is studied via pressure dependent measurements. Our results constitute an important step towards realizing integrated optomechanical circuits employing hBN.
\end{abstract}

\maketitle

Integration of nanoscale photonic and mechanical resonators into cavity optomechanical devices \cite{ref:kippenberg2007com, ref:aspelmeyer2014co} has enabled fundamental discoveries and applications spanning quantum information science \cite{ref:chan2011lcn, ref:teufel2011scm, ref:schliesser2009rsc, ref:cohen2015pci, ref:purdy2017qcrt, ref:sudhir2017qco, ref:hong2017hbt}, sensing \cite{ref:anetsberger2009nco, ref:srinivasan2011oti, ref:kim2013nto, ref:forstner2012com, ref:yu2015cot, ref:schilling2016nfi, ref:wu2017not}, and optical signal processing \cite{ref:fiore2011soi, ref:hill2012cow, ref:liu2013eit, ref:fang2017gnr, ref:ruesink2018ocm}. Key to these breakthroughs is the ability of nanoscale cavity optomechanical devices to enhance the interaction between light and motion of mechanical resonators, and to provide sensitive transduction of this interaction via its effect on the  response of narrow  cavity optical resonances.

A natural application of cavity optomechanics is the study and manipulation of the mechanical properties of 2D materials \cite{ref:cole2015efo, ref:song2014gor}, whose intrinsically nanoscale dimensions can make observing and controlling their mechanical motion challenging. It is these same properties that make 2D materials attractive for many applications of nanomechanics whose performance can be enhanced with a decrease in resonator mass \cite{ref:aspelmeyer2014co}, for example molecule detection  \cite{ref:ekinci2004uli, ref:gil2015hfn, ref:roy2018ims}. Recently, hexagonal boron nitride (hBN) has attracted considerable attention as a promising platform to study nanophotonic effects in 2D materials \cite{ref:tran2016quantum, ref:xia2014two}. hBN is a hyperbolic material that supports propagation of phonon polaritons, assisting in confining light to the deep sub-wavelength regime \cite{ref:caldwell2014sub, ref:dai2014tunable}, and potentially enabling enhanced optical forces \cite{ref:he2012gto}. Such effects have yet to be explored in optomechanics, as hBN photonic devices are in early stages of development and have previously not been studied within the context of optomechanics. Moreover, hBN is a promising 2D van der Waals material as it hosts ultra bright single photon sources that operate at room temperature \cite{ref:tran2016quantum, ref:exarhos2017optical, ref:jungwirth2016temperature}, whose spectral properties are sensitive to mechanical strain. This makes hBN optomechanical devices an attractive system to implement experiments in the emerging field of spin-optomechanics, as discussed in \cite{ref:abdi2017spin}. Within the field of quantum optomechanics, 2D materials hold tremendous potential as they enable the creation of low mass devices, with associated large zero point motion compared to those fabricated from more conventional dielectric films. hBN's high transparency from visible to infrared wavelengths, which stems from its 6 eV bandgap \cite{ref:cassabois2016hexagonal}, combined with its potential for strong optical confinement via its aforementioned hyperbolic properties, make it an ideal 2D material for implementing such quantum devices.

\begin{figure*}[t]
\centering
\includegraphics[width=\linewidth]{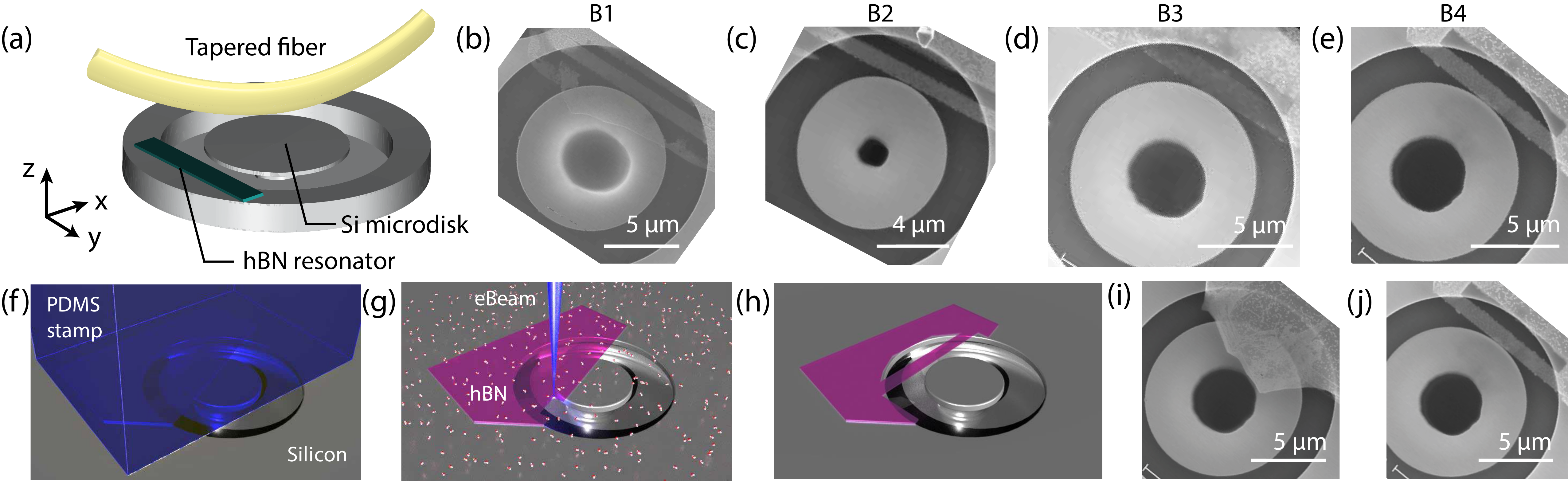}
\caption{(a) Schematic diagram, and (b-e)  SEM images of the hBN-Si devices studied in this work, labelled B1-B4. Note that the (b) SEM image of device B1 was taken during the fabrication process, before undesired hBN was removed from the microdisk top surface. (f-h) Device fabrication process.  (f) A suitable hBN flake is transferred using a PDMS stamp to a position adjacent to a pre-fabricated Si microdisk. (g) Maskless, direct EBIE step in water vapour for hBN etching, which enables the selective fabrication of (h) a nanobeam next to the microdisk. (i,j) SEM images of device B4 before and after EBIE etching.}
\label{fig:schematic}
\end{figure*}

In this work, we take an initial step towards these applications by demonstrating the first study of a cavity optomechanical system employing hBN nanomechanical resonators, and show that the mechanical properties of these nanoscale devices can be sensitively probed using this system. hBN can be exfoliated into flakes with thickness as low as a single monolayer, and their mechanical properties can be studied when these flakes are partially suspended \cite{ref:li2016atomically, ref:falin2017mechanical, ref:zheng2017hexagonal}. Recently, a novel, top-down hBN patterning technique combining reactive ion etching (RIE) and focused electron beam induced etching (EBIE) was developed \cite{ref:elbadawi2016electron}, allowing the direct nanoscale patterning of photonic crystal cavities made entirely from monolithic hBN \cite{ref:KIm2018pcc}. Here, we utilize this technique to create nanomechanical resonators from hBN that are integrated with silicon (Si) nanophotonic devices to realize a cavity optomechanical system. Thermally driven motion of the hBN mechanical resonator is read-out via its interaction with a high optical quality factor ($Q_\text{o} \sim 10^4 - 10^5$) Si microdisk, allowing observation of mechanical resonances with quality factor $Q_\text{m} > 10^3$ -- among the highest reported for 2D materials at room temperature \cite{ref:castellanos2015mechanics}, and exceeding previously reported values for hBN resonators \cite{ref:li2016atomically, ref:cartamil2017mechanical, ref:falin2017mechanical, ref:zheng2017hexagonal}.

The hBN-Si cavity optomechanical systems studied here are shown in Figure \ref{fig:schematic} together with an overview of the fabrication process. The system consists of an hBN nanobeam suspended adjacent to 220 nm thick Si microdisk optical cavity. hBN-Si microdisks were fabricated on two different silicon chips for the studies presented here, with device B1 using an 8.4$\,\mu\text{m}$ diameter silicon microdisk, and devices B2 - B4 using 11.6$\,\mu\text{m}$ diameter microdisks. The dimensions of the hBN nanobeams fabricated in this work varied depending on the fabrication details described below, with the measured dimensions of each B1 - B5 given in the Supporting Information. Note that the reported hBN thickness is different for each device, and although not directly measured here, was estimated during the fabrication process to be less than approximately 50 nm for all of the devices, which is consistent with what is predicted from the nanomechanics measurements presented below. The hBN nanobeams are positioned between $< 50$ nm and $160~\text{nm}$ from the microdisk edge depending on the device, as shown in Fig.\ 1(b -- e) and in the Supporting Information. These gaps are small enough to observe fluctuations in the nanobeam position via the microdisk optical response.  In the following, we assume that the hBN has a refractive index $n\sim1.8$ at the telecommunication band operating wavelength of the microdisk \cite{ref:cassabois2016hexagonal}.

The microdisk fabrication process follows Ref.\  \cite{ref:borselli2005brs}, where they are first patterned in the Si layer of a silicon-on-insulator chip using electron beam lithography and reactive ion etching, followed by HF undercutting to remove the underlying $3~\mu\text{m}$ of SiO$_2$ until the microdisks are supported by thin SiO$_2$ pedestals. The process for integrating an hBN mechanical resonator with the microdisk is summarized in Figs.\ \ref{fig:schematic}(f - h). The electron beam lithography pattern used to define the microdisk creates a ring shaped trench surrounding the device over which an hBN nanobeam can be suspended. The first step in creating the nanobeam is to transfer an hBN flake with homogeneous thickness and high quality adjacent to the pre-fabricated microdisk using a dry PDMS stamp transfer technique (Fig.\ \ref{fig:schematic}(f)). EBIE is then used to define a nanomechanical resonator  separated by a small gap from the Si microdisk. During this process, a focused electron beam in the presence of water vapour induces selective chemical reactions with hBN, which leads to localized volatilization (Fig.\ \ref{fig:schematic}(g)) \cite{ref:elbadawi2016electron}. Since EBIE is entirely chemically driven, this process does not destroy the adjacent Si microdisk as evidenced by the SEM images in \ref{fig:schematic}(b-e), and the high optical $Q_\text{o}$ of the device presented below. Additionally, EBIE has not been observed to induce crystalline damage to the hBN nanobeam, compared to other methods such as focused ion beam milling \cite{ref:KIm2018pcc}. For further illustration of the fabrication process, SEM images of a device before and after EBIE are shown in Fig.\ \ref{fig:schematic}(i,j). For detailed before and after fabrication images of all the devices, see the Supporting Information. An advantage of this in-situ patterning technique is that it allows the gap between the nanobeam and the microdisk to be finely adjusted. This is of critical importance, as large gaps between hBN resonator and Si microdisk prevent observation of optomechanical coupling. While the serial nature of the etching process is not immediately scalable, it is well suited for studying this hBN-Si hybrid system as it is non invasive to the underlying silicon structure. It also allows prototyping devices, which are in turn used here for initial characterization of the nanomechanical properties of the hBN nanobeams with high sensitivity.

\begin{figure*}[!ht]
\centering
\includegraphics[width=\linewidth]{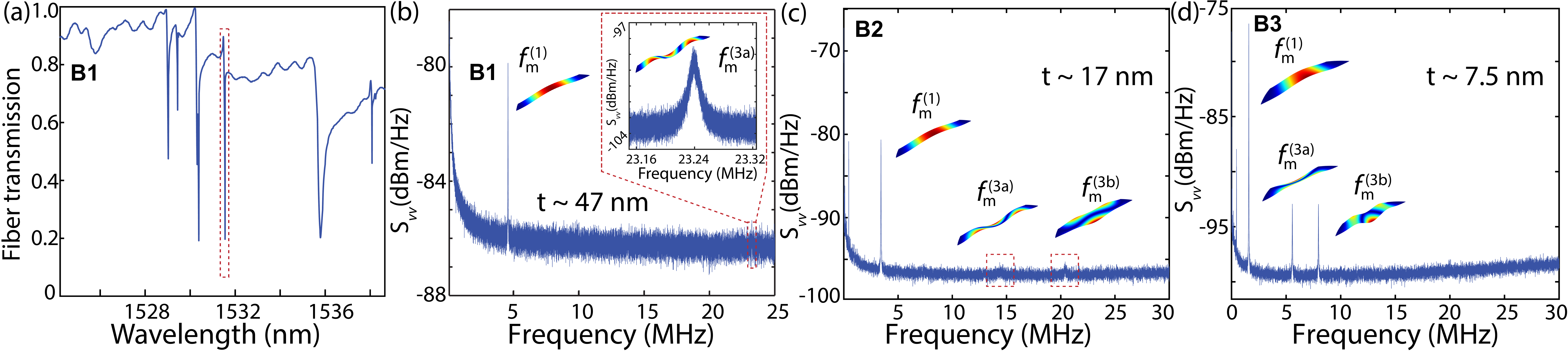}
\caption{(a) Normalized fiber taper transmission as a function of input laser wavelength when the fiber is in contact with the Si microdisk. (b-d) Power spectral density of the photodetected fiber taper output intensity  when the input laser is near-resonance with the optical mode. Peaks corresponding to optomechanically transduced thermal motion of a mechanical mode of the nanobeam are visible, and are labeled by their simulated displacement profiles for device B1, B2, and B3, respectively. Note that technical noise peaks from the apparatus were removed as described in the Supporting Information, and a probable fiber mode is present in (c,d) at $\sim$ 450kHz. Inset: zoomed-in view of the $f_\text{m}^{(3a)}$ mode of hBN for device B1.}
\label{fig:spectra}
\end{figure*}

Measurement of the optomechanical properties of the hBN-Si system was achieved using a dimpled optical fiber taper to couple light into and out of the Si microdisk \cite{ref:michael2007oft}. In order to reduce damping of the hBN nanobeam's motion, measurements were performed in a vacuum chamber (base pressure, $P \sim 2 \times 10^{-5}~\text{Torr}$), inside of which nanopositioners (Attocube) were used to position the device and the optical fiber taper. Figure \ref{fig:spectra}(a) shows a typical transmission spectrum of the fiber taper when it placed in contact with the microdisk, evanescently coupling input light to it via a tunable laser (New Focus TLB-6700). Sharp dips in the transmission correspond to coupling to whispering gallery mode resonances of the microdisk, while broad low amplitude dips in transmission are related to wavelength dependent variations in the fiber taper transmission and laser output.

Mechanical motion of the nanobeam was probed by fixing the input laser wavelength $\lambda$ within a high-$Q_\text{o}$ resonance, and measuring the electronic power spectral density $S_{vv}(f)$ of the photoreceiver output (New Focus 1811) as in Ref.\ \cite{ref:wu2014ddo}.
Typical spectra for multiple devices are shown in Fig.\ \ref{fig:spectra}(b-d). For frequencies, $f < 1$  MHz, these spectra are dominated by $1/f$ noise. However, peaks in the spectrum associated with thermally driven motion of normal mechanical modes of the hBN nanobeams are identifiable at higher frequencies.  For example, in the case of the B1 device spectrum in Fig.\ \ref{fig:spectra}(b), peaks with resonance frequencies $f_\text{m}^{(1)} = 4.6~\text{MHz}$ and $f_\text{m}^{(3a)} = 23~\text{MHz}$ are visible. Based on Lorentzian fits, these resonances are observed to have linewidths corresponding to $Q_\text{m} = 260$ and $1100$, respectively.  The high-$Q_\text{o}$ optical resonance near $1531.7~\text{nm}$  used in this measurement is highlighted in Fig.\ \ref{fig:spectra}(a), and found to have $Q_\text{o} \sim 88,000$ (see below) and dominant electric field polarization along the radial direction (TE --like). To improve the visibility of these nanobeam related peaks, after the study of device B1, devices B2 -- B5 were fabricated with nominally smaller nanobeam-microdisk gaps. The smallest fabricated gaps, e.g.\ for device B3, was found to be below the 50 nm resolution of the SEM images taken during device fabrication. Higher resolution images were not taken to avoid damaging the hBN nanobeam.
The spectra in Figs.\ \ref{fig:spectra}(c) and (d) of devices B2 and B3, respectively, have a larger signal to noise as well as an additional third visible resonance. Table 1 summarizes the measured resonance frequencies and quality factors for devices B1 -- B5.


\bgroup
\def\arraystretch{2}
\begin{table*}[t]
\begin{center}
\rowcolors{1}{}{tablecolor}
\begin{tabular}{*{9}{>{\centering\arraybackslash}p{0.102\linewidth}}}
\hline
  Device & $t$ (nm) & $f_m^{(1)}$ (MHz) & $f_m^{(3a)}$ (MHz) & $f_m^{(3b)}$ (MHz) & $Q_\text{m}^{(1)}$ &$Q_\text{m}^{(3a)}$&$Q_\text{m}^{(3b)}$ & $f_m^{(3b)}/f_m^{(3a)}$ \\ 
  B1 & 47  & 4.6 (4.6) & 23.24 (22.8) & - & 260 & 1100 & - & - \\ 
  B2 & 17 & 3.42 (3.4) & 14.45 (14.3) & 20.4 (20.9) & 215 & 214 & 930 & 1.4 \\ 
  B3 & 7.5 & 1.58 (1.6) & 5.56 (5.55) & 7.96 (8.1) & 278 & 224 & 298 & 1.4\\ 
  B4 & 18.5 & 3.54 (3.5) & 12.99 (13.0) & 16.8 (17.2) & 678 & 630 & 1098 & 1.3 \\ 
  B5 & 51 & 4.6 (4.6) & 15.7 (15.5) & 20.8 (20.6) & 657 & 770 & 684 & 1.3 \\ 
\end{tabular}
 \caption{Frequencies of the measured (simulated) hBN nanobeam mechanical resonances, as well as the measured quality factors of the first and third order modes, and the nanobeam thicknesses used in the simulations. Note that simulations of B3, B4 and B5 included a small vertical curvature in the beam, as described in the Supporting Information. }
\label{table:ratio}
\end{center}
\end{table*}
\egroup

To better understand the nanomechanical nature of these resonances, finite element simulations (COMSOL) were used to calculate the normal mode spectrum of the nanobeams under study. These were then compared with the observed resonances, as shown in Table 1. This process is complicated somewhat by uncertainty in the thickness of each nanobeam stemming from lack of angled high resolution SEM images, and unwanted etching of the hBN nanobeam from EBIE, as described in the Supporting Information. Assuming a Young's modulus of 865 GPa \cite{ref:falin2017mechanical}, for a nanobeam thickness of 31 nm for device B1 we find good agreement between the  observed resonances and predicted $f_\text{m}$ of 4.59 MHz and 23.24 MHz for the 1st ($f_\text{m}^{(1)}$) and 3rd ($f_\text{m}^{(3a)}$) order vertical modes of the nanobeam, respectively. The simulated displacement profiles of these modes are shown in Fig.\ \ref{fig:spectra}(b). Other studies have reported lower Young's modulus, e.g.\ 392 GPa \cite{ref:zheng2017hexagonal}, which could be related to differences in material quality or the presence of internal compressive stress. Using this lower Young's modulus in our simulations results in a predicted hBN thickness of 47 nm for device B1. This lower value of Young's modulus is used throughout the remaining studies as the larger thickness is most consistent with the expected dimensions. Atomic force microscope measurements, combined with high resolution  angled SEM images would provide more accurate characterization of the hBN thickness, and while not employed here to avoid risk of damaging the devices, will be useful in future studies to help reduce this ambiguity.

The $2^\text{nd}$ order vertical mode of device B1 is not observed due to the odd symmetry of its displacement profile with respect to the $x$-axis (defined in Fig.\ \ref{fig:schematic}(a)) and the even symmetry of the microdisk mode intensity about this axis. As a result, the optomechanical coupling coefficient, $G$, which predicts the shift in the frequency of the cavity mode for a given mechanical displacement $z$, vanishes, resulting in nominally no cavity optomechanical transduction of the motion of this mode \cite{ref:schilling2016nfi}. The lowest order horizontal mechanical mode is predicted to have $f_\text{m} = 131.87$ MHz for device B1, and was also not observed. This could be explained by its lower thermally driven amplitude ($\propto 1/f_\text{m}^3$) owing to its high frequency. Note that this effect is also responsible for the low observed amplitude of the $f_\text{m}^{(3a)}$ peak.

As shown in Table 1, devices B2 and B3 are predicted by comparison to simulation to have thickness of 17 nm and 7.5 nm, respectively. The small thicknesses of B3 is in qualitative agreement with its SEM image in Fig.\ \ref{fig:schematic}(d), in which the nanobeam is highly transparent. The additional 3rd resonance at frequency $f_\text{m}^{(3b)}$ of devices B2 and B3 was also predicted by simulations, and originates from  the asymmetry along the $y$-axis of the hBN nanobeams studied here. The two resulting 3rd-order vertical modes, $f_\text{m}^{(3a)}$ and $f_\text{m}^{(3b)}$, have maximum displacement on one each of the two long edges of the beam, as shown in Figs.\ \ref{fig:spectra}(c-d)). This asymmetry arises due to the circular shape of the window over which the hBN nanobeams are clamped by van der Waals forces. The ratio $f_\text{m}^{(3b)}/f_\text{m}^{(3a)}$ was observed to be relatively constant across devices, as expected from analytic theory \cite{ref:brand2015rmf}. Measurements of device B1 were unable to resolve the $f_\text{m}^{(3b)}$ mode, likely due to its relatively high frequency of 32.3 MHz and the poor signal to noise for this device. In order to find agreement between simulation and measured frequencies for device B3 it was necessary to include a 3 nm vertical curvature at the center of the beam. This curvature was not necessary for devices B1 and B2, and a larger curvature was necessary for B4 and B5 (see Supporting Information). This indicates that compressive stress may be present in these devices. Devices B4 and B5 also displayed additional resonances not explained by simulations and requiring further investigation in future work (see Supporting Information). Mechanical $Q_\text{m}$ of the measured fundamental and higher order modes varied between 215 and 1100 for the measured devices, with no clear correlation between $Q_\text{m}$ and device thickness. As hBN is a van der Waals material it may experience strong slipping between layers upon straining\cite{ref:kobayashi2017sal}, which could in turn lead to variation in the observed mechanical properties. In this work only primarily macroscopic strain effects were considered, however this system could provide a useful platform for studying these nanoscopic effects.

\begin{figure}
\centering
\includegraphics[width=\linewidth]{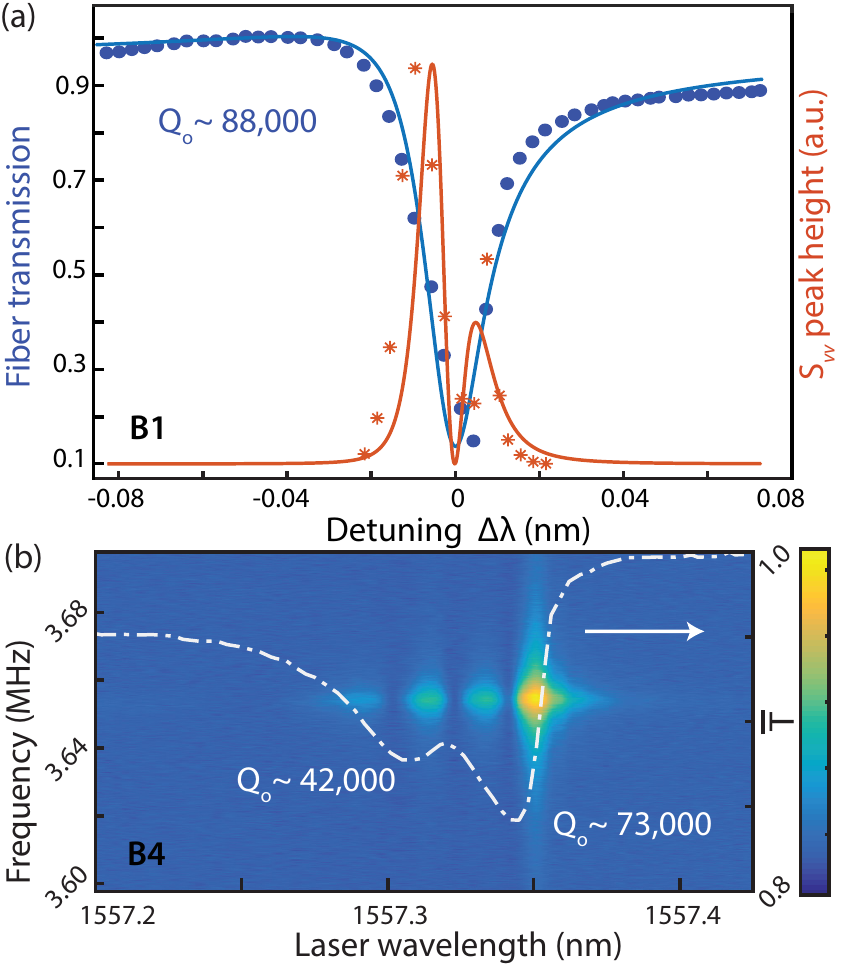}
\caption{(a) Wavelength dependence of the peak height of $S_{vv}(f)$ for the $f_\text{m}^{(1)}$ mode of device B1 when the laser is tuned across the microdisk cavity resonance, together with the corresponding fiber taper transmission. The fit to the $S_{vv}(f)$ peak height (orange line) is obtained from the slope of the fit to the Fano lineshape of the fiber taper transmission (blue line). A small amount of thermo-optic nonlinearity in the microdisk response not accounted for in the fit is responsible for the slight asymmetry of the lineshape. (b) Spectrograph $S_{vv}(f_\text{m}^{(1)};\lambda)$ for device B4 when the laser is scanned across a microdisk doublet cavity resonance, overlayed with the  corresponding fiber taper transmission in white. The range of the colorbar scale is -115 dBm/Hz to -95 dBm/Hz.}
\label{fig:detuning}
\end{figure}

To better confirm that the peaks are related to optomechanically transduced hBN motion, their amplitudes were measured as a function of input laser detuning $\Delta\lambda = \lambda - \lambda_\text{o}$ from the microdisk resonance wavelength $\lambda_\text{o}$. This is shown in Fig.\ \ref{fig:detuning}(a) for the $f_\text{m}^{(1)}$ mode of device B1, from which we see that the peak amplitude  follows the slope of the cavity optical response: $S_{vv}(f_\text{m}^{(1)};\lambda) \propto |dT(\lambda)/d\lambda|^2$, and that it is maximum when the laser is near the point of maximum slope. This behaviour is consistent with the sideband unresolved system studied here exhibiting predominantly dispersive optomechanical coupling between the hBN and the microdisk. Dominantly dispersive coupling is possible due to the low optical absorption of hBN at the operating wavelength, in contrast to the dissipative  optomechanical coupling observed in highly absorbing graphene cavity optomechanical systems \cite{ref:cole2015efo}, indicative of hBN's suitability for cavity optomechanics. Note that the asymmetry in $S_{vv}(\Delta\lambda)$ is well predicted from the asymmetry in the lineshape of the microdisk resonance, which arises from interference effects related to higher order modes of the fiber taper waveguide \cite{ref:wu2014ddo} as well as small thermal instability in the microdisk \cite{ref:barclay2005nrs}.

To further illustrate that the optomechanical coupling demonstrated here is primarily dispersive in nature, a spectrograph showing the wavelength dependence of the power spectral density was measured for the $f_\text{m}^{(1)}$ mode of device B4, as shown in Fig.\ \ref{fig:detuning}(b). In this measurement an optical doublet mode corresponding to nearly degenerate standing wave modes created by backscattering in the microdisk \cite{ref:borselli2005brs} was used, with $Q_\text{o}\sim 42,000$ for the symmetric mode and $Q_\text{o}\sim 73,000$ for anti-symmetric mode. As for the previous device, we observed that $S_{vv}(f_\text{m}^{(1)};\lambda) \propto |dT(\lambda)/d\lambda|^2$, with maxima when the laser wavelength is near the points of maximum slope for each mode of the doublet.

Several additional observations were used to confirm that the measurements described above are not related to normal modes of the microdisk itself. The lowest frequency microdisk mode with non-zero nominal optomechanical coupling is the radial breathing mode \cite{ref:mitchell2014cog, ref:sun2012hqs}, whose frequency is orders of magnitude higher than resonances observed here. Lower frequency mechanical modes of the microdisk can in principle have small non-zero optomechanical coupling resulting from asymmetries in device geometry. However, COMSOL simulations (see Supporting Information) of the B1 microdisk indicate that its lowest frequency modes are at 2.5 MHz, 9.9 MHz, 14.4 MHz and 33.4 MHz, which do not correspond to the observed resonances. Finally, all measurements were carried out with the fiber taper in contact with the microdisk, which should damp its mechanical motion significantly.

\begin{figure}
\centering
\includegraphics[width=\linewidth]{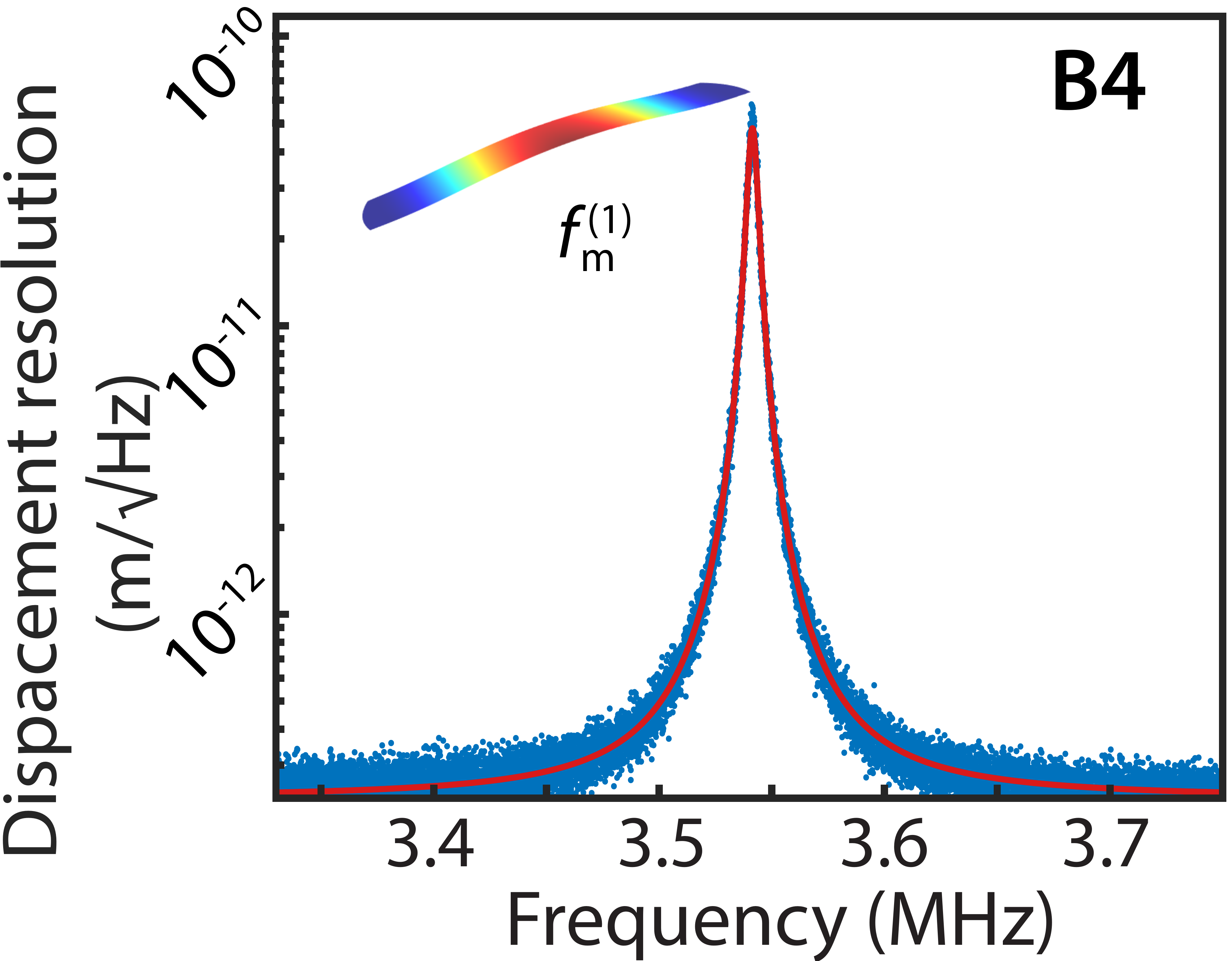}
\caption{Displacement resolution for device B4, exhibiting the largest SNR observed of all devices, with its Lorentzian fit, ($f_\text{m}^{(1)} \sim 3.54 $MHz, $Q_\text{m}\sim 678$), corresponding to $s_{xx}^\text{min} =160$ fm/Hz$^{1/2}$.}
\label{fig:snr}
\end{figure}

The measurement resolution of the cavity optomechanical system can be extracted from the measured signal to noise ratio $\alpha$ of each thermomechanically driven peak using the expression $s_{xx}^\text{min} = \sqrt{\frac{4k_\text{B}TQ_\text{m}/(m_\text{eff}\Omega_\text{m}^3)}{\alpha}}$, where $\Omega_\text{m} = 2\pi f_\text{m}$, $k_\text{B}$ is Boltzmann's contstant and $T = 300~K$ is the device temperature \cite{ref:hauer2013gpt, ref:wu2014ddo}. Here $\alpha$ is defined by the ratio of the resonance peak height $S_{vv}(f_\text{m})$ to the measurement noise floor, and  $m_\text{eff}$ is the predicted effective mass of the  normal mode, as defined in Ref.\ \cite{ref:aspelmeyer2014co}. From  COMSOL simulations we find $m_\text{eff} = 160$ fg, $83.9$ fg, and $50.8$ fg for the $f_\text{m}^{(1)}$, $f_\text{m}^{(3a)}$ and $f_\text{m}^{(3b)}$ modes of device B4, respectively, assuming an hBN density of 2100 kg/m$^3$ \cite{ref:cartamil2017mechanical, ref:rumyantsev2001pas}. Note that this anomalous decrease in $m_\text{eff}$ with mode order is related to the nanobeam's non right-angle clamping points and  resulting trapezoidal shape. From the data in Fig.\ \ref{fig:snr}, which was obtained with $\lambda$ tuned to maximize the signal, we find  $\alpha = 243.4$, corresponding to $s_{xx}^\text{min} = 160$ fm/Hz$^{1/2}$. This is smaller than comparable graphene cavity optomechanical systems \cite{ref:cole2015efo} but larger than what has been achieved in optomized SiN nanobeam cavity optomechanical device \cite{ref:schneider2016soc}. Sensitivity could be further improved through operation at lower temperature, enhancements of $Q_o$ to the $10^6$ level observed by Borselli et al.\ \cite{ref:borselli2005brs}, and as discussed below, increased $G$. In addition, an input optical power, $P_\text{in}\sim 93~\mu\text{W}$ was used for these measurements, limited by the presence of thermally--induced optical shifts from two photon absorption in the Si microdisk \cite{ref:barclay2005nrs}. As $\alpha$ typically increases with $P_\text{in}$ \cite{ref:wu2014ddo}, improved measurement resolution could be achieved using a microdisk fabricated from a material which does not suffer from nonlinear absorption at $\lambda\sim1550$ nm and is compatible with the fabrication process, such as GaP\cite{ref:mitchell2014cog}.

The optomechanical coupling affects the measurement resolution through $S_{vv} \propto G^{2}$, and can be optimized through the positioning of the nanobeam relative to the microdisk \cite{ref:schilling2016nfi}. To assess $G$ in our system, we implemented the calibration technique (see Supporting Information) from Refs.\ \cite{ref:gorodetksy2010dvo, ref:schneider2016soc} which adds a known phase modulation to the input laser that is then transduced by the cavity into an optical intensity modulation. From the measured tone  height relative to the area under nanomechanical resonance peak, we extracted $G/2\pi = 0.4$ MHz/nm for device B1. This is smaller than, but in reasonable agreement with,  $G/2\pi = 0.6$ MHz/nm predicted from a perturbation theory calculations of the dependence of $\lambda_\text{o}$ on the nanobeam height  \cite{ref:schilling2016nfi}.  This calculation was performed by evaluating the overlap of the microdisk mode intensity with the nanobeam dielectric, where  finite difference time domain simulations (MEEP \cite{ref:oskooi2010mff}) were used to determine the  azimuthaly symmetric field (fundamental TE-like mode, azimuthal number $m=55$, travelling wave field distribution) of the unperturbed microdisk. This calculation does not take into account the local field correction to the microdisk field by the nanobeam \cite{ref:johnson2002ptm}, which may account for the discrepancy. Significant uncertainty can also arise from variation in the gap between the hBN nanobeam and the microdisk, as $G$ depends exponentially on this parameter \cite{ref:schilling2016nfi}.

After characterizing device B1, devices B2 - B5 were fabricated with a smaller gap in order to increase $G$. While this is largely borne out by the larger signal to noise ratio of the spectra measured here, care must be taken to disentangle differences in $G$ from variations in $Q_o$, optical mode order, and microdisk and nanobeam dimensions. A systematic study of the influence of gap size on $G$, and its effect on $Q_o$, is not undertaken here but is required in future work to fully optimize this system. However, the good agreement between the measured and predicted value of $G$ for B1 indicates that executing this optimization numerically is likely to yield valuable insight.

\begin{figure}
\centering
\includegraphics[width=\linewidth]{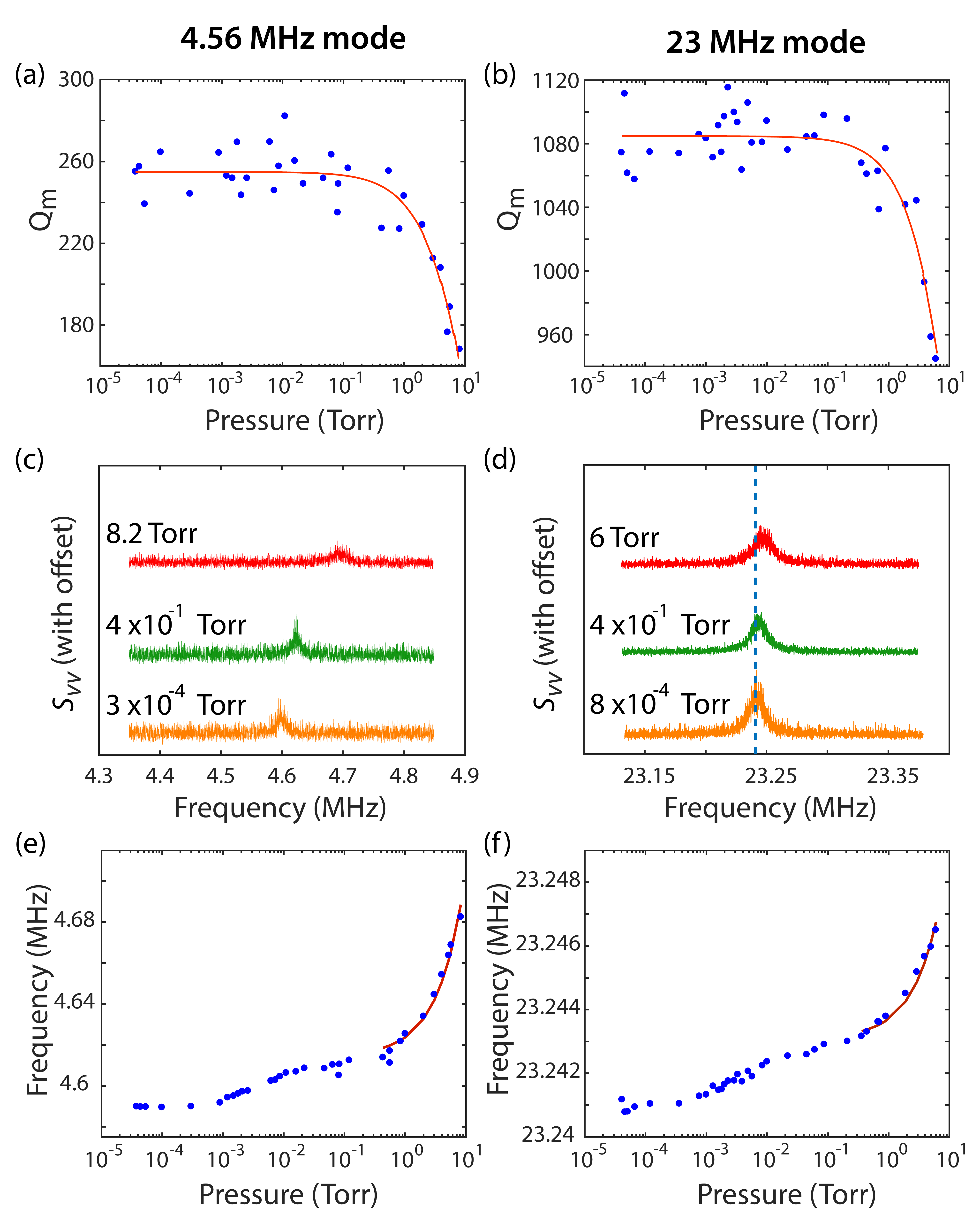}
\caption{ (a,b) Dependence of $Q_\text{m}$ on pressure for the $f_\text{m}^{(1)}$ and $f_\text{m}^{(3a)}$ modes of device B1, respectively. The red lines show  fits following the free molecular flow damping model described in the text. (c,d) Power spectral density of the photodetected fiber taper output intensity at different pressures for the $f_\text{m}^{(1)}$ and $f_\text{m}^{(3a)}$ modes, respectively. The dashed line in (d) is a guide for the eye. (e,f)  Dependence of mechanical frequency on vacuum pressure for the $f_\text{m}^{(1)}$ and $f_\text{m}^{(3a)}$ modes, respectively. The red lines are fits based on the squeeze film spring effect.}
\label{fig:pressure}
\end{figure}

The  properties of the hBN nanomechanical resonator were further probed by studying its dynamics as a function of vacuum pressure. Figures \ref{fig:pressure}(a) and \ref{fig:pressure}(b) show the observed dependence of $Q_\text{m}$ for the $f_\text{m}^{(1)}$ and $f_\text{m}^{(3a)}$ modes of device B1, respectively. In each case, $Q_\text{m}(P)$ is unaffected by pressure for $P < 10^{-1}~\text{Torr}$, taking a constant value $Q_\text{vac}$ in this pressure range. This indicates that air damping is not limiting $Q_\text{m}$ for the measurements reported above. At higher pressure, as shown by the fits in Figs.\ \ref{fig:pressure}(a-b), the pressure dependence of $Q$ was well modeled by $1/Q_\text{m} = 1/Q_\text{vac} + 1/Q_\text{fmf}(P)$, where $1/Q_\text{fmf} \propto P$ describes the influence of free molecular flow damping \cite{ref:blom1992dqf, ref:verbridge2008sfd, ref:lee2014ada}. Note that the relatively similar pressures at which $Q_\text{m}$ of the $f_\text{m}^{(1)}$ and $f_\text{m}^{(3a)}$ modes begin to degrade is in agreement with the analytically predicted dependence of $Q_\text{fmf}$ on mode frequency \cite{ref:verbridge2008sfd, ref:lee2014ada}.

The mechanical frequencies of the hBN nanomechanical resonators were also observed to change with $P$, as shown in Figs.\ \ref{fig:pressure}(c) and \ref{fig:pressure}(d). This pressure dependent shift can be caused by several mechanisms. For $P \gtrapprox 1~\text{Torr}$, the shift follows $\Delta f_\text{m}^2(P) \propto P$. This behaviour is consistent with a squeeze film spring effect observed in nanomechanical resonators when the oscillation period is smaller than the response time of the gas molecules trapped between the resonator and the substrate \cite{ref:dolleman2015gsf}. This effect requires that the air mean free path is sufficiently small for molecules to fill the gap between the resonator and the substrate. However, at $P \sim 1~\text{Torr}$ the mean free path of air is $\sim 40~\mu\text{m}$, which is larger than the maximum dimension ($\sim 16~\mu\text{m}$) of the aperture defined by the gap between the nanobeam and substrate in device B1. This suggests that the vacuum chamber pressure in the vicinity of the device may be larger than measured by the pressure gauge, which is located near the turbo pump used to evacuate the chamber. Note that the chamber used here is a Montana Instruments Nanoscale Workstation nominally designed for low temperature operation.

Below $P \sim 0.1~\text{Torr}$ the frequencies of both modes also increase with pressure, although with a different pressure dependence. Similar behavior was observed by Lee et al.\, where it was ascribed to stress induced by the local pressure differential experienced by the nanobeam caused due to restriction of air molecules from entering the volume between the nanobeam and the underlying silicon chip \cite{ref:lee2014ada}, though other studies suggest that a steady state pressure should be reached \cite{ref:dolleman2016ggo}. Finally, note that optical power dependent measurements of $f_\text{m}$ confirmed that local photothermal heating was not influencing the results reported here, as described in the Supporting Information.

Note that for the measurement in Fig.\ \ref{fig:pressure}(c,d) a different optical mode, at 1489.7 nm ($Q_o\sim87,600$), was used than in other measurements of device B1. This optical mode had improved transduction for the $f_\text{m}^{(3a)}$ mechanical mode at the expense of smaller transduction for the first order mechanical mode, allowing measurement of small frequency shifts of the third order mode.  This change in relative transduction strength is possibly related to differences in optical mode order, which affects the optical field overlap with the nanobeam.

To conclude, we have experimentally demonstrated an hBN cavity optomechanical system with displacement sensitivity $s_{xx}^\text{min} = 160$ fm/Hz$^{1/2}$, and have shown that hBN nanobeams support mechanical resonances with $Q_\text{m}$ exceeding 1000 at 23 MHz frequency in room temperature vacuum conditions. Our results are the first step towards integration of hBN as vital component of an integrated quantum optomechanical systems. Within the context of spin-optomechanics, future studies will examine whether optomechanical control of hBN nanomechanical resonators can be used to modify the emission properties of hBN emitters. Future research will also examine whether novel or enhanced optical forces resulting from the hyperbolic nature of hBN can be demonstrated. While the hBN-Si system presented here would be unsuitable for collection of single photon emission from hBN or studying hBN's optomechanical properties in the lower Restrahlen band due to silicon's optical absorption at those wavelength, this Si system could be used for studying optomechanics in the upper Restrahlen band. Furthermore, the flexibility of the fabrication process described in this work would allow integration of the hBN structures with microdisks fabricated from materials who do not suffer absorption at these wavelengths, such as GaP \cite{ref:mitchell2014cog}, which would allow the study of the single photon emitters present in hBN.

In the immediate future, experiments will be focused on optimizing the design of the hBN nanomechanical resonator and its integration with the optical cavity, fabricating optomechanical crystals \cite{ref:eichenfield2009oc} directly from hBN and implementation of other fully integrated hBN cavity optomechanical devices, and measurement of hBN nanomechanical properties at cryogenic temperatures. Such experiments would provide additional insight into the nanomechanical properties of hBN resonators and their interaction with optical fields, and allow development sub-fm/$\sqrt{\text{Hz}}$ readout of hBN resonator motion \cite{ref:schneider2016soc}.

\textbf{Acknowledgments} We thank the referees and R.\ Dolleman for their helpful comments. Financial support from the Australian Research council (via DP180100077, LP170100150), the Asian Office of Aerospace Research
and Development grant FA2386-17-1-4064, the Office of Naval Research Global under
grant number N62909-18-1-2025, the Natural Science and Engineering Research Council, and the National Research Council are gratefully acknowledged.

%

\clearpage

\onecolumngrid

\setcounter{equation}{0}
\setcounter{figure}{0}
\setcounter{section}{0}
\setcounter{subsection}{0}
\setcounter{table}{0}
\setcounter{page}{1}
\makeatletter
\renewcommand{\theequation}{S\arabic{equation}}
\renewcommand{\thetable}{S\arabic{table}}
\renewcommand{\thefigure}{S\arabic{figure}}
\renewcommand{ \citenumfont}[1]{S#1}
\renewcommand{\bibnumfmt}[1]{[S#1]}

\section*{Supplementary Information}
\subsection{Fabrication \& estimated device dimensions}

\begin{figure*}[h]
\centering
\includegraphics[width=\linewidth]{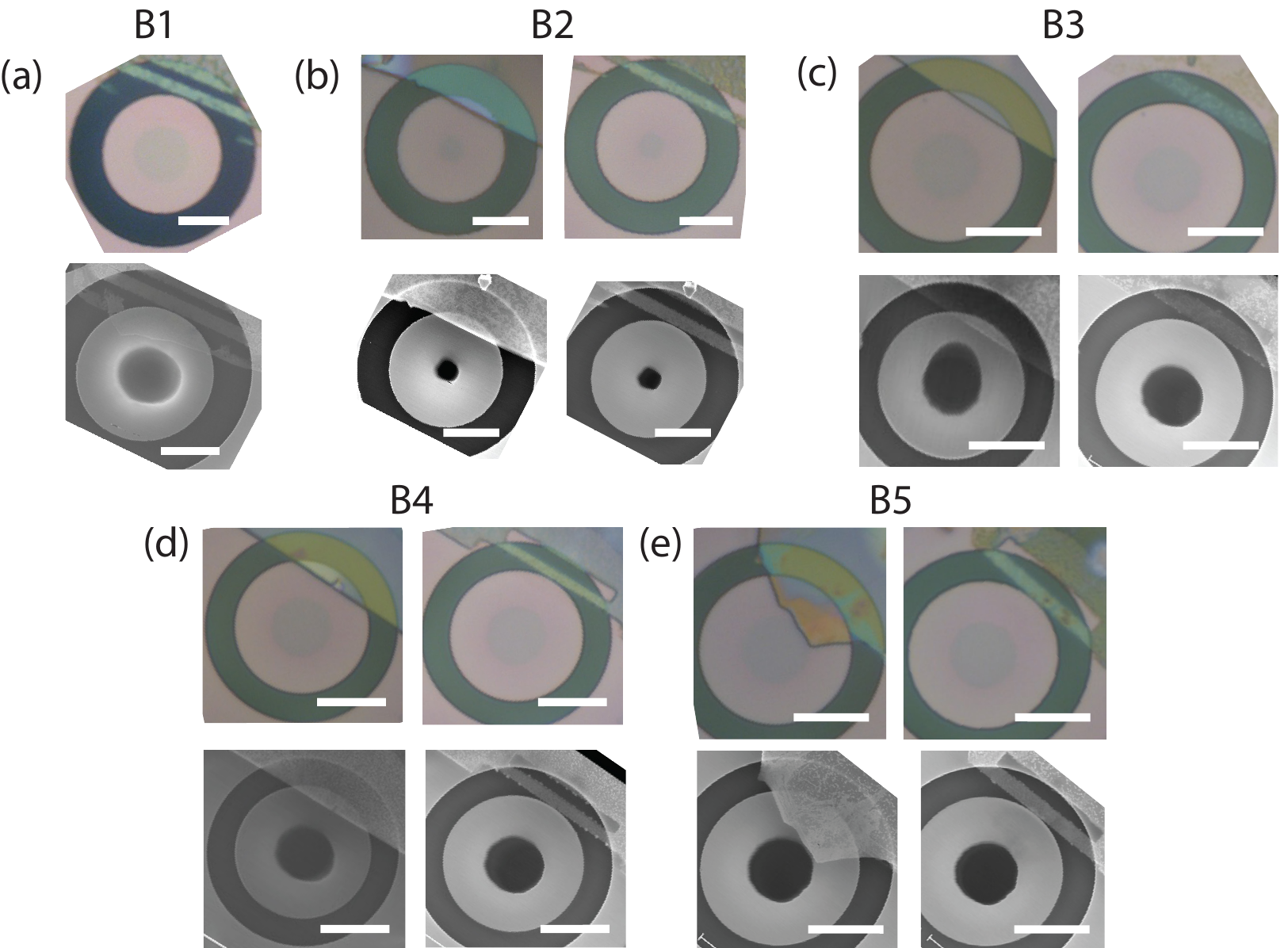}
\caption{Optical microscope and SEM images of the hBN beams before, during, and after EBIE. (a) hBN beam defined by EBIE where hBN flake is still present on the microdisk.(b,c) before and after EBIE patterning of the hBN beam and removal of the remaining hBN flake from the microdisk. The scale bars correspond to (a,c-e) 5 $\mu$m  (b) 4 $\mu$m.}
\label{fig:before_after}
\end{figure*}

Figure\ \ref{fig:before_after} shows optical and scanning electron micrographs (SEM) before and after electron beam induced etching (EBIE) process for all of the devices studied in this work. We note that the colour change of the hBN flake in the optical microscope images is associated with undesired etching in areas that are not directly exposed, as well as positioning the area of interest in the center of the electron beam, which will etch the surface layer \cite{ref:supp_elbadawi2016electron}. This can potentially be bypassed by operating at lower electron beam energy, giving a smaller radius of backscattered electrons, which at this stage is a technical limitation.

Table\ \ref{table:ratio} gives the estimated dimensions for all devices studied in this work. The hBN beam length, $l$, and width, $w$ were measured from SEM images, while the thickness $t$ was estimated by tuning it to match the simulated and measured mode frequencies, $f_\text{m}$. Note that for devices B3-B5, a vertical curvature with height ($h_\text{v}$) had to be introduced to find good agreement with the measured mechanical modes, indicating that compressive stress may be present in these devices.

\bgroup
\def\arraystretch{2}
\begin{table}[ht]
\begin{center}
\rowcolors{1}{}{tablecolor}
\begin{tabular}{*{8}{>{\centering\arraybackslash}p{0.097\linewidth}}}
\hline
  Device & $l$ ($\mu$m) & $w$ ($\mu$m)& $t$ (nm) & $h_\text{v}$ (nm) & $f_\text{m}^{(1)}$ (MHz) & $f_\text{m}^{(3a)}$ (MHz) & $f_\text{m}^{(3b)}$ (MHz) \\
  B1 & 13.5 & 1.1 & 47 & 0 & 4.6 & 22.8 & 32.3 \\
  B2 & 10.04 & 0.92 & 17 & 0 & 3.4 & 14.3 & 20.9\\
  B3 & 12.7 & 1.6 & 7.5 & 3 & 1.6 & 5.55 & 8.1\\
  B4 & 11.4 & 1.15 & 18.5 & 9 & 3.5 & 13.0 & 17.2\\
  B5 & 13.4 & 1.08 & 51 & 20 & 4.6 & 15.5& 20.6\\
\end{tabular}
 \caption{COMSOL estimated thickness ($t$) and vertical curvature height, $h_\text{v}$, of hBN nanobeams with Young's modulus of 392 GPa, density of 2100 kg/m$^3$ and simulated values of the first and third order modes for all the devices. $l$ is the length of longest edge of the beam.}
\label{table:ratio}
\end{center}
\end{table}
\egroup

\subsection{Microdisk mechanical normal modes}

\begin{figure*}[h]
\centering
\includegraphics[width=\linewidth]{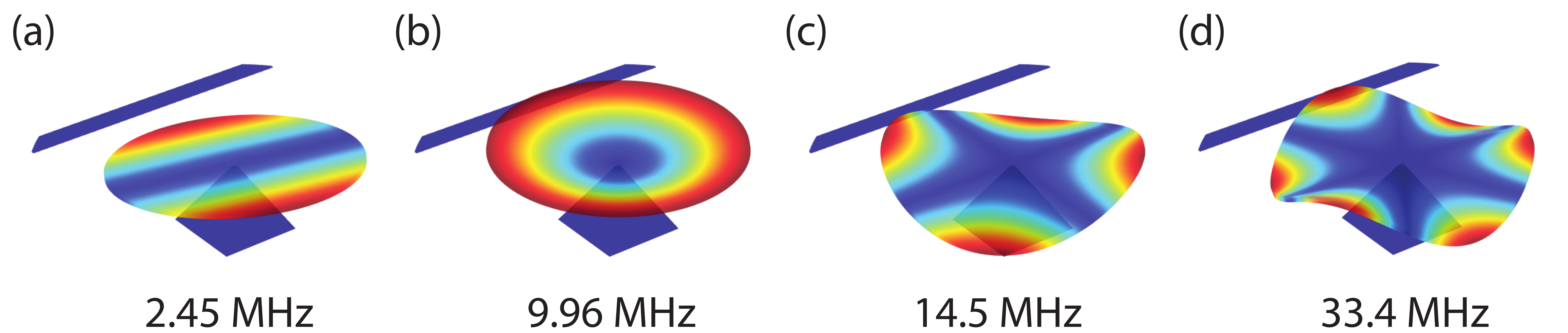}
\caption{Displacement profile and corresponding mechanical frequency for microdisk in device B1 (diameter = 11.6 $\mu$m, thickness = 220 nm). The measured frequencies of the hBN nanobeam modes for device B1 are 4.6 MHz and 23.24MHz.}
\label{fig:devices}
\end{figure*}

To rule out the presence of mechanical modes of the microdisk in the measured power spectral density, $S_{vv}(f)$, the microdisk normal modes were simulated in COMSOL, using a disk thickness of 220 nm defined by the silicon on insulator material, and SEM measured diameter of 11.6 $\mu$m. The lowest order modes which could have small non-zero optomechanical coupling resulting from asymmetries in device geometry are shown in Figure\ \ref{fig:devices}. As they are spectrally separated from the expected hBN nanobeam modes, and the fiber taper induced damping of its motion should be significant, we conclude that we do not observe these modes in our measured spectra.

\subsection{Power dependence of $S_{vv}(f)$}

\begin{figure*}[h]
\centering
\includegraphics[width=\linewidth]{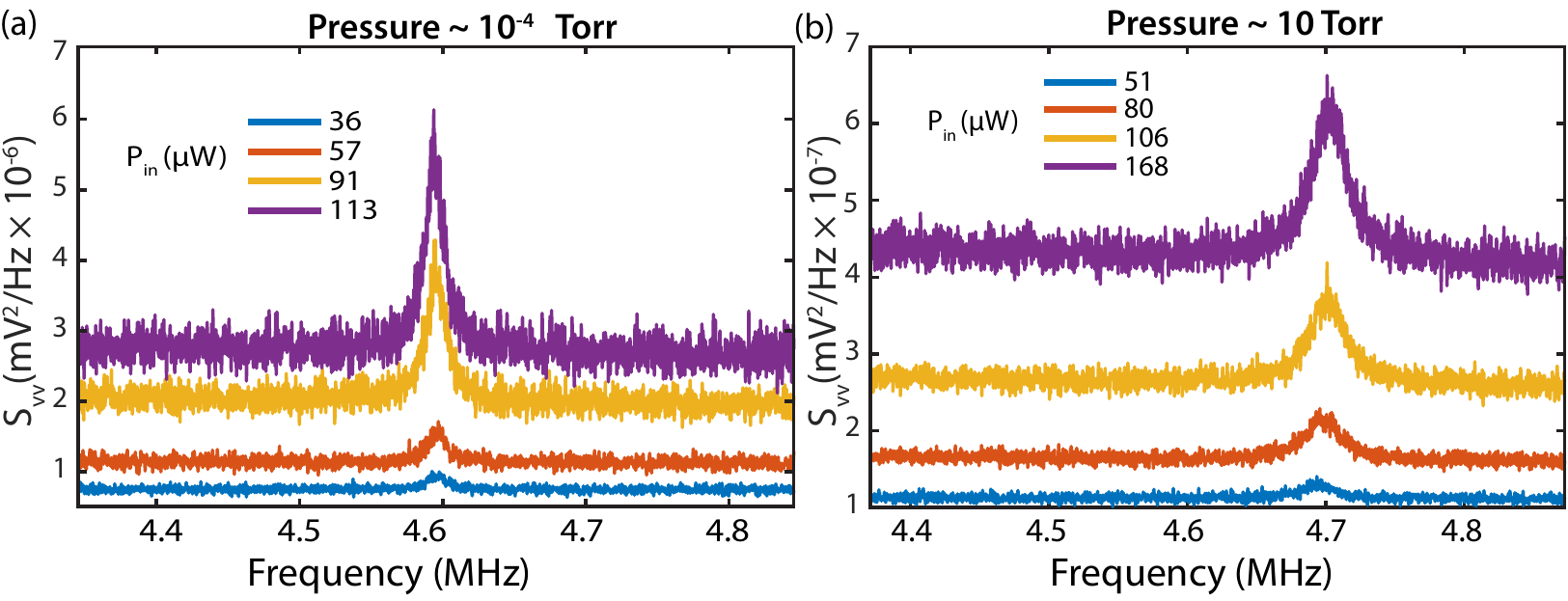}
\caption{(a,b) $S_{vv}(f)$ for various optical input power, $P_\text{in}$, for the $f_\text{m}^{(1)}$ mode of device B1 at chamber pressures of $\sim10^{-4}$ Torr and $\sim10$ Torr, respectively.}
\label{fig:power}
\end{figure*}

To rule out the shift in mechanical frequency due to local photothermal heating we studied the dependence of the measured power spectral density, $S_{vv}(f)$, of the $f_\text{m}^{(1)}$ mode on input optical power, $P_{in}$. As shown in Figures\ \ref{fig:power}(a) and \ref{fig:power}(b) the frequency of the $f_\text{m}^{(1)}$ mode of device B1 was observed to remain constant with increasing $P_{in}$, for chamber pressures of $\sim10^{-4}$ and $\sim10$ Torr, respectively.

\newpage
\subsection{Optomechanical coupling coefficient calibration}

\begin{figure*}[h]
\centering
\includegraphics[width=0.5\linewidth]{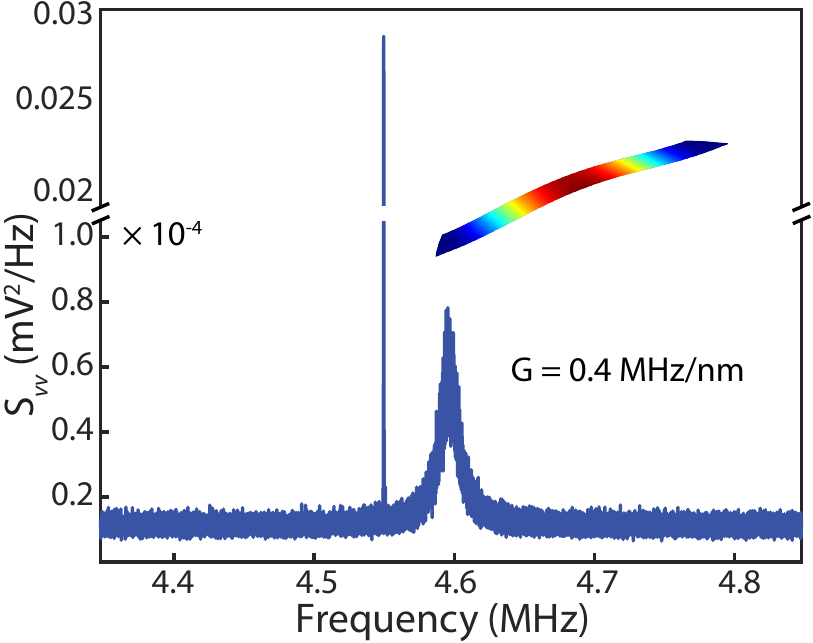}
\caption{$G$ was determined by comparing the thermomechanical cavity frequency fluctuation with a calibration tone, which is generated by phase modulating the input laser field. The right peak is the Lorentzian mechanical resonance, and the left peak is the phase modulated calibration tone.}
\label{fig:calibration}
\end{figure*}

To assess the optomechanical coupling coefficient, $G$\cite{ref:supp_aspelmeyer2014co}, in our system, we implemented the calibration technique developed in Ref.\ \cite{ref:supp_gorodetksy2010dvo} which adds a known phase modulation to the input laser that is then transduced by the cavity into an optical intensity modulation. From the measured area under the modulation tone relative to the area under nanomechanical resonance peak shown in Figure\ \ref{fig:calibration} we extracted $G/2\pi = 0.4$ MHz/nm for device B1. This phase tone was created by passing the input laser through a electro--optic phase modulator (EOSpace) modulated at a frequency close to $f_\text{m}^{(1)}$.

\subsection{Additional spectra and background noise removal}

\begin{figure}[h]
\centering
\includegraphics[width=0.56\linewidth]{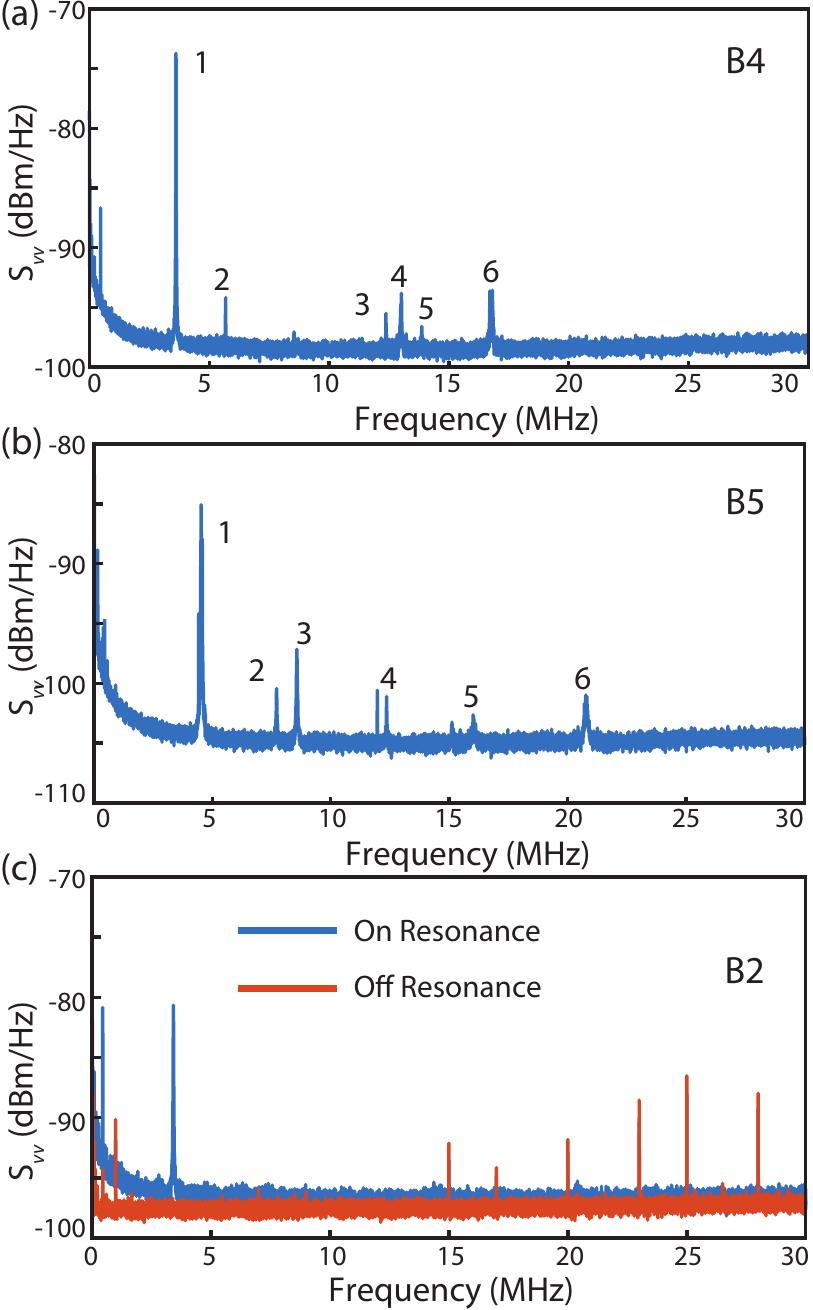}
\caption{$S_{vv}(f)$ spectra of the photodetected fiber taper output intensity when on-resonance for the hBN beam. Modes marked (a) 1 and 4,6 and (b) 1 and 5,6 are the first and third order modes of the hBN nanobeam respectively. (c) Example of $S_{vv}(f)$ when the laser is tuned within the optical resonance and off resonance. These background spectra were used to remove noise peaks from the $S_{vv}(f)$ in (a,b) and in Figure 2 of the main text.}
\label{fig:additional_device}
\end{figure}

Wide band measurements of $S_{vv}(f)$ for device B4, and device B5 (not discussed in the main text) is shown in Figure\ \ref{fig:additional_device}(a,b) where, as in the main text, these measurements were carried out in vacuum (base pressure, P $\sim 3\times10^{-3}$ Torr), with the same methods described in the main text. For devices B4 and B5, we observed hybridization of certain modes which would exhibit multiple peaks in a narrow--band about the expected center frequency. A potential reason for this hybridization could be asymmetry present in the nanobeam due to thickness variations from EBIE and non-rectangular clamping areas at each end of the nanobeam and will be studied in future work.

To remove spurious noise features in $S_{vv}(f)$ resulting from electronic noise in our measurement apparatus a $S_{vv}(f)$ spectra was taken when the laser was placed off-- resonance with the optical mode of the Si microdisk, while the fiber was parked. By comparing this off--resonance $S_{vv}(f)$ spectra to that when on--resonance, noise features were identified and removed from the $S_{vv}(f)$ spectra presented in the main text and and Figure\ \ref{fig:additional_device}(a,b).

\end{document}